\documentclass[12pt]{article}

\usepackage[paper=a4paper, margin=2cm]{geometry}
\usepackage[russian,english]{babel}
\usepackage[cp1251]{inputenc}
\usepackage{amsmath}

\usepackage[logos]{amsbib}

\begin{document}
%%%%%%%%%%%%%%%%%%%%%%%%%%%%%%%%%%%%%%%%%%%%%%%%%%%%%%%%%%%%%%%%%%%%%%%%%%%%%%%%%%%%

\date{}
\renewcommand{\refname}{References}

\author{A.~Yu.~Samarin%, samarinay@yahoo.com
}
\title{Non-mechanical nature of the wave function collapse }

\maketitle

\centerline{Samara Technical State University, 443100 Samara, Russia}
\centerline{}

\abstract{The quantum particle is considered as an indivisible continuous medium transformed into a mass points as the result of the collapse in the form of the space localization. The wave function transformation of the quantum particle is described by the integral evolution operator with the kernel in the form of a path integral. It is shown that this approach allows considering not only Schr{\"{o}}dinger's evolution, but also the collapse phenomenon. The nature of the collapse is the nonlocal transformation of the quantum particle internal structure and it is not connected with its mechanical motion. Probably, this fact allows for instantaneous signalling using the collapse without violating of the relativistic requirements.


%%%%%%%%%%%%%%%%%%%%%%%%%%%%%%%%%%%%%%%%%%%%%%%%%%%%%%%%%%%%%%%%%%%%%%%%%%%%%%%%%%%%

\begin {center} \textbf{1.Introduction: the root of the problem}\\
\end {center}
 The wave function of the quantum particle cannot be interpreted "as giving somehow the density of the stuff of which the world is made"~\cite[p.227]{bib:Aa}, unless to solve the problem of the wave packet spreading ~\cite{bib:Ba,bib:Bb,bib:Bd}\footnote{Such an approach would let us to attribute the uncertainty generating quantum stochasticity to the properties of the measuring instrument, that allows exclude a truly nonepictemic character from quantum probabilities.}. This problem is solved  by itself, if there is the possibility of the quantum particle space localization during any infinitesimal time interval no matter of the spacial region size, where the the wave function of the particle differs from zero before the localization.

 The nonlocality of the reduction phenomenon is logical consequence of quantum mechanics laws~\cite{bib:Be,bib:Bf}. Experimental confirmation of this statement~\cite{bib:Bi,bib:Bj,bib:Bg} gives a possibility to consider the wave function collapse as a real transformation of extended substance. The fact that in some of these experiments the observer influence on their results is impossible in principle ~\cite{bib:Bk,bib:Bl} forces us to seek the collapse cause in the substance itself.

 As the object of analysis, for simplicity, we consider a quantum particle. In accordance with the idea mentioned above, the quantum particle is an indivisible quantum object, that is transformed into a mass point as the result of the collapse.

 \begin {center} \textbf{2.The law of dynamics}\\
\end {center}

 Since the space localization leads to the singularity of the wave function, the integral form of quantum evolution law~\cite{bib:Bm},~\cite[p.57]{bib:Bn} is more suitable for the analysis of the collapse than the differential one.

 Let us consider the system consisting of the studied quantum particle (further a particle-object) and the particles of the measuring instrument (hereinafter active particles) interacting with this particle-object. For simplicity, one-dimensional motion is only considered. By $x$ denote the coordinate of the particle-object, by $X$ --- the generalized coordinates of the active particles collection, by $i$ -- the number of the generalized coordinate, by $dX=\prod_{i=1}\limits^{N}dX^{i}$ -- the volume element of the configuration space of the active particles. The evolution of the system is described by the  integral wave equation~\cite{bib:Bm,bib:Bn} in the form
 \begin{equation}\label{eq:math:ex1}
 {\Psi_{t_{2}}(x_{2},X_{2})
   = \int K_{t_{2},t_{1}}(x_{2}X_{2},x_{1},X_{2})\Psi_{t_{1}}(x_{1}X_{1})dx_{1}dX_{1}}
\end{equation}
where $\Psi_{t_{2}}(x_{2},X_{2}) $, $\Psi_{t_{1}}(x_{1},X_{1}) $ are the wave functions of the particle at the time $t_{2} $ and the initial time $t_{1}<t_{2} $, respectively. The subscript of the spatial variable denotes the time.  Denote by $\Gamma $ a virtual path in the configuration space of the system. Then, for the kernel of the integral evolution operator $K_{t_{2},t_{1}}(x_{2},x_{1}) $ in accordance with~\cite{bib:Bo} we have
\begin{equation}\label{eq:math:ex2}
{K_{t_{2},t_{1}}(x_{2},x_{1})=\int \exp{\frac{i}{\hbar}S_{1,2}[\Gamma]}\,[d\Gamma]}.
\end{equation}

 Integral wave equation~\eqref{eq:math:ex1} describes the change with time of the field of the wave function values. Without discussing the physical sense of this quantity, we assert that it has a material basis. If we consider the quantum particle as a continuous medium, then this basis is an individual particle. The specific features of the quantum continuum are considered in ref.~\cite{bib:Bp}, where the integral wave equation is considered as the Eulerian method of the mechanical motion description. Thus, the cause of the wave function transformation in the Schr{\"{o}}dinger evolution is the mechanical motion of individual particles. But it is not the only method of the wave function transformation, which the integral wave equation provides. Really, the instantaneous infinite jump of the potential energy entering in the action functional in expression~\eqref{eq:math:ex2} in a local region of space result in the increase in the measure of the set of the virtual paths passing through the corresponding region of space at an appropriate time. In accordance with expressions~\eqref{eq:math:ex1} and~\eqref{eq:math:ex2} this generates a transformation of the wave function. The Schr{\"{o}}dinger equation does not consider this possibility. When it is derived from the integral wave equation~\cite{bib:Bn,bib:Br}, the potential energy is supposed invariable for an infinitesimal time interval.

\begin {center} \textbf{3. The wave function collapse initiation}\\
\end {center}
In order to understand the possible cause of such jump, let us consider the action functional in~\eqref{eq:math:ex2}. By $U_{q}(x,X)$, denote the interaction potential energy of the active particles with the particle-object; by $U_{A}(X)$ denote the potential energy of the active particles in the external field; by $U(x,t)$ -- the potential energy of the particle-object in the external field. Then for the action functional we have
\begin{equation}
 S_{1,2}[\Gamma]=\int\limits_{t_{1}}^{t_{2}}\biggl(T^{\Gamma}(\dot{X},\dot{x})-U_{q}(x,X)-U_{A}(X)h(t-t_3)-U(x,t)\biggl) \,dt.
\label{eq:math:ex3}
\end{equation}
In this expression $T^{\Gamma}(\dot{X},\dot{x}) $ denotes the kinetic energy of all the particles of the system; $h(t-t_3) $ is the Heaviside function. The Heaviside function expresses the initiation of a macroscopic process (hereinafter a registering process) in the measuring instrument at the time $t_3$. Formally, this means the following. When the energy of the active particles increases above a threshold limit value(due to the interaction with the quantum particle), the region of configuration space accessible for them  immediately changes; this generates the transformation of the virtual paths set; the potential energy for the paths of this new set has a macroscopic value.

Suppose that the registering process takes place in a small local spatial domain as this occures when the particle coordinate is measured. Denote by $\Omega^{I}$ the region of the configuration space corresponding to this domain and by $\Omega^{II}$ -- the rest configuration space. The initiation of the registering process in this domain is expressed mathematically by the potential energy jump of the corresponding active particles in~\eqref{eq:math:ex3}. Let us consider the state of the system just after this jump. At this time the potential energy $U_{A}(X)$ having a macroscopic value considerably exceeds all other terms in expression~\eqref{eq:math:ex3}. Denote by $\varepsilon=t_{4}-t_{3}$ the time interval, after which the action functionals on all the paths of all microscopic processes become negligible small compared with the value $U_{A}\varepsilon$. The kinetic energy of the registering process can be neglected, because it is an infinitesimal quantity compared with the potential energy within this time interval. By $\Gamma$ denote the virtual path of the active particles, by $\gamma$ -- the path of the particle-object. Then, taking into account~\eqref{eq:math:ex3}, for the transition amplitude~\eqref{eq:math:ex2} we obtain
\begin{equation}
 K_{t_{4},t_{3}}^{I}(x_{4},X_{4},x_{3},X_{3})=\exp\biggl(-{\frac{i}{\hbar}U_{a}\varepsilon}\biggl)\Biggl(\int\limits_{\Omega^{I}}\exp{\frac{i}{\hbar}s_{3,4}[\gamma]}\,[d\gamma]\Biggl)\delta(X_{4}-X_{3}),
\label {eq:math:ex4}
\end{equation}
for $\Omega^{I} $ and
\begin{eqnarray}\label{eq:math:ex5}
K_{t_{4},t_{3}}^{II}(x_{4},X_{4},X_{4},x_{3},X_{3})=\\
 =\int\limits_{\Omega^{II}} \exp{\frac{i}{\hbar}s_{3,4}[\gamma]}\Biggl(\int\exp{\frac{i}{\hbar}S_{3,4}[\Gamma]} \,[d\Gamma]\Biggl)\,[d\Gamma], \nonumber
\end{eqnarray}
for $\Omega^{II} $. In order to ascertain the form of the wave function at the time $t_{4}$, it is useful to transform the expressions for these amplitudes into a real form.

\begin {center} \textbf{4. The result of the  potential energy jump }\\
\end {center}

The quantum path integral can be reduce to the real form of the path integral having the Wiener measure~\cite{ex16}.  In order to make this transformation, it is necessary to represent the time variable in the complex form  $t=\tau\exp\bigl(-i\varphi\bigr)$ and consider the transition amplitude for $\varphi=-\frac{\pi}{2}$ i.e. $t=-i\tau $~\cite{ex14}.  This mathematical manipulation reverses the chain of events. It can be used when we consider the  Schr{\"{o}}dinger evolution (because the classical mechanical motion along the virtual path is reversible). However, it cannot be used in the case of the irreversible quantum jump~\cite{bib:Bs}.In the last case we have to consider the real path integral for $t=i\tau$. The measure of the path integral can be extended analytically on this part of the complex time plane~\cite{bib:Bt}. After such transformation for the transition amplitudes~\eqref{eq:math:ex4} and ~\eqref{eq:math:ex5}, we obtain
\begin{equation}\label {eq:math:ex6}
 K_{\tau_{4},\tau_{3}}^{I}(x_{4},X_{4},x_{3},X_{3})=\exp{\frac{1}{\hbar}U_{a}\varepsilon}\Biggl(\int\limits_{\Omega^{I}}\exp{\frac{i}{\hbar}s_{3,4}[\gamma]}\,[d\gamma]\Biggl)\delta(X_{4}-X_{3}),
\end{equation}
for $\Omega^{I} $ and
\begin{eqnarray}\label{eq:math:ex7}
K_{t_{4},t_{3}}^{II}(x_{4},X_{4},x_{3},X_{3})=\\
 =\int\limits_{\Omega^{II}} \exp{\biggl(-\frac{1}{\hbar}s_{3,4}[\gamma]\biggl)}\Biggl(\int\exp{\biggl(-\frac{1}{\hbar}S_{3,4}[\Gamma]\biggl)} \,[d\Gamma]\Biggl)\,[d\Gamma], \nonumber
\end{eqnarray}
for $\Omega^{II} $. In the last expressions the quantities $s[\gamma], S[\Gamma] $ are the action functionals written for modulus $\tau$ of the complex time variable.

The conventional normalization of the wave function expressing the quantum particle integrity have to be conserve in the localization process. Therefore after the collapse we obtain
\begin{equation*}
\int\int\Biggl|\Psi^{I}_{t_{4}}(x,X)+\Psi^{II}_{t_{4}}(x,X)\Biggl|^{2}\,dx\,dX=1.
\end{equation*}
For imaginary time we have
\begin{equation*}
\int\int\Biggl|\Psi^{I}_{\tau_{4}}(x_{4},X_{4},)+\Psi^{II}_{\tau_{4}}(x_{4},X{4})\Biggl|^{2}\,dx_{4}\,dX_{4}=1,
\end{equation*}
where
\begin{eqnarray}
\Psi^{I}_{\tau_{4}}(x_{4},X_{4})\nonumber\\=\int\int K_{\tau_{4},\tau_{3}}^{I}(x_{4},X_{4},x_{3},X_{3})\Psi_{\tau_{3}}(x_{3},X_{3})\,dx_{3}\,dX_{3};\nonumber
\end{eqnarray}
\begin{eqnarray}
\Psi^{II}_{\tau_{4}}(x_{4},X_{4})\nonumber\\=\int\int K_{\tau_{4},\tau_{3}}^{II}(x_{4},X_{4},x_{3},X_{3})\Psi_{\tau_{3}}(x_{3},X_{3})\,dx_{3}\,dX_{3}.\nonumber
\end{eqnarray}
The amplitude $K_{\tau_{4},\tau_{3}}^{I}$ considerably exceeds the amplitude $K_{\tau_{4},\tau_{3}}^{II}$ due to the macroscopic order of magnitude of the exponent in this term. This means that
\begin{equation*}
\int\int\Biggl|\Psi^{I}_{\tau_{4}}(x_{4},X_{4})\Biggl|^{2}\,dx_{4}\,dX_{4}\approx 1.
\end{equation*}
Let us suppose that this domain is infinitesimal(i.e corresponding volume of space occupied by the active particles taking part in the registering process is infinitesimal) and the interaction radius of the quantum particle with the active particle is infinitesimal, too. Let $Y^{I}$ be the space coordinate of this volume (such situation takes place when particle coordinate is measured). Then, all path $\gamma$ at the time $t_{4}$  pass pass trough this point, and at the time $\tau_{4}$ we have a localized state
\begin{displaymath}
\Psi_{\tau_{4}}(x_{4},X_{4})=\delta\left(x_{4}-Y^{I}(\tau_{4})\right)\Phi(X_{4}).
\end{displaymath}
The wave function of the active particles after the collapse has the form:
\begin{displaymath}
\Phi_{\tau_{4}}(X_{4})=\delta\left(X^{k}_{4}-X_{I}^{k}(\tau_{4})\right)\Phi_{\tau_{4}}(X_{4}^{1},...,X_{4}^{k-1},X_{4}^{k+1},...,X_{4}^{N}),
\end{displaymath}
where $X^{k}_{4}$ is the generalized coordinates of the active particles describing the registering process (after its initiation); $X_{I}^{k}(\tau_{4})$ -- the values of this generalized coordinates of the classical registering process at the time $\tau_{4}$. Then for the real time we have
\begin{displaymath}
\Psi_{t_{4}}(x_{4},X_{4})=\delta\left(x_{4}-Y^{I}(t_{4})\right)\Phi_{t_{4}}(X_{4}).
\end{displaymath}
Thus, after the collapse the entangled wave function of the system is transformed into the product of the wave function of the the measuring instrument and the wave function particle-object having the form of delta function. At the time $t_{4}$ the interaction of the particle-object and the measuring instrument is instantly terminated everywhere with the exception of the domain $\Omega^{I}$, and the particle-object is visualized macroscopically as a single point.
\begin {center} \textbf{5. Discussion}\\
\end {center}

The change of the measure of the virtual paths set under the collapse
is generated by the infinite large value of the local temporal derivative of the potential energy corresponding to these paths, i.e. the collapse is not caused by the mechanical motion of individual particles. It is not a form of of the  mechanical motion of the quantum particle, but the
transformation of the internal structure of the integral extended object (in the sense of the physical identity of all the parts). Such transformation have to be a nonlocal process, that does not violate the relativistic requirements attributing to the mechanical motion.

In the case considered, we have the nonlinear transformation of the wave function described by integral wave equation~\eqref{eq:math:ex1}. The density matrix of the statistical ensemble of the quantum particles is transformed nonlinearly, too. At the same time, equation~\eqref{eq:math:ex1}. The density matrix of the statistical ensemble of the quantum particles is transformed nonlinearly, too. At the same time, equation\eqref{eq:math:ex1} is inherently deterministic (the collapse stochasticity is caused by the statistical straggling of the measuring instrument parameters). In accordance with the theorem proved in~\cite{bib:D}, it means that the collapse phenomenon enables the information transfer. This allows for faster than light communication~\cite{bib:Bx,bib:By}.  This problem, maybe, is seeming. Really, the cause of the difficulties is in identification of the communication with the transfer of a material signal in space. Then, if this transfer is determined by the spatio-temporal motion of a physical object, we must operate within the relativistic requirements and the problem is real. However, it is possible, that this way of communication is not unique. Since the collapse phenomenon is not a form of the spatio-temporal motion, it is not the object of the consideration of relativistic theory and, therefore, it can be used for the instantaneous communication without any violation of its requirements.

\vfill\eject

\end{document}